\documentclass[aps,prb,twocolumn,preprintnumbers,amsmath,amssymb]{revtex4-1}

\usepackage{times}
\usepackage{graphicx}
\usepackage{dcolumn}
\usepackage{bm}
\usepackage{natbib}
\usepackage{hyperref}
\usepackage{amssymb}
\usepackage{amsmath}
\usepackage{SIunits}

\usepackage{float}
\usepackage{color}

\hypersetup{backref,pdfpagemode=UseOutlines,colorlinks=true}%

\begin{document}
\title{Optical properties of an ensemble of G-centers in silicon}

\author{C. Beaufils,$^{1}$ W. Redjem,$^{1}$ E. Rousseau,$^{1}$ V. Jacques,$^{1}$ A. Yu. Kuznetsov,$^{2}$ C. Raynaud,$^{3}$ C. Voisin,$^{3}$ A. Benali,$^{4}$ T. Herzig,$^{5}$ S. Pezzagna,$^{5}$ J. Meijer,$^{5}$ M. Abbarchi,$^{4,\dag}$ G. Cassabois$^{1,\ast}$}

\affiliation{$^{1}$Laboratoire Charles Coulomb (L2C), Univ. Montpellier, CNRS, Montpellier, France\\
$^{2}$Department of Physics, University of Oslo, NO-0316 Oslo, Norway\\
$^{3}$Laboratoire Pierre Aigrain, Ecole Normale Sup\'{e}rieure, Universit\'{e} Paris Diderot, UPMC, CNRS UMR8551, 24 rue Lhomond, 75005 Paris, France\\
$^{4}$ CNRS, Aix-Marseille Universit\'{e}, Centrale Marseille, IM2NP, UMR 7334, Campus de St. J\'{e}r\^{o}me, 13397 Marseille, France\\
$^{5}$Department of Nuclear Solid State Physics, Felix-Bloch Institute for solid-state physics, Universitat Leipzig, Linn\'{e}stra\ss{}e 5, 04103 Leipzig, Germany}

\date{\today}

\begin{abstract}
We addressed the carrier dynamics in so-called G-centers in silicon (consisting of substitutional-interstitial carbon pairs interacting with interstitial silicons) obtained via ion implantation into a silicon-on-insulator wafer. For this point defect in silicon emitting in the telecommunication wavelength range, we unravel the recombination dynamics by time-resolved photoluminescence spectroscopy. More specifically, we performed detailed photoluminescence experiments as a function of excitation energy, incident power, irradiation fluence and temperature in order to study the impact of radiative and non-radiative recombination channels on the spectrum, yield and lifetime of G-centers. The sharp line emitting at 969 meV ($\sim$1280 nm) and the broad asymmetric sideband developing at lower energy share the same recombination dynamics as shown by time-resolved experiments performed selectively on each spectral component. This feature accounts for the common origin of the two emission bands which are unambiguously attributed to the zero-phonon line and to the corresponding phonon sideband. In the framework of the Huang-Rhys theory with non-perturbative calculations, we reach an estimation of 1.6$\pm$0.1 $\angstrom$ for the spatial extension of the electronic wave function in the G-center. The radiative recombination time measured at low temperature lies in the 6 ns-range. The estimation of both radiative and non-radiative recombination rates as a function of temperature further demonstrate a constant radiative lifetime. Finally, although G-centers are shallow levels in silicon, we find a value of the Debye-Waller factor comparable to deep levels in wide-bandgap materials. Our results point out the potential of G-centers as a solid-state light source to be integrated into opto-electronic devices within a common silicon platform.         
\end{abstract}
\maketitle

\section{INTRODUCTION}\label{sec:INTRO}

Semiconductor materials are at the heart of the technology in our knowledge-based society. The most important one is silicon, which has become a cornerstone in the electronics and photovoltaics industries. However, the indirect nature of the bandgap in silicon is a severe drawback for opto-electronics applications. In spite of the potentials of silicon-based photonic devices, the fundamental issue of light emission remains a challenge. For that reason, a number of solutions have been explored, e.g. alloying silicon and germanium, doping and strain engineering \cite{LOC1997,SUE2000,SUN2009,CHE2009,CLA2011,SAN2011,JAI2012,CAM2012,SUE2013}. For instance, in silicon-germanium nanostructures, such as quantum wells, nanocrystals and nanowires, the quantum confinement of carriers leads to enhanced absorption and spatially direct transitions in the visible and infrared range \cite{LOC1996,DEH2000,PAR2001,OSS2003,GRY2016,GRY2016b,SCH2017}.

An alternative viable strategy towards the integration of optical and electronic devices on the same silicon-based platform relies on extrinsic impurities embedded in the host semiconductor matrix. Relaxing the need of a direct bandgap, the presence of extrinsic centers acting as deep levels allows for optical emission. A plethora of impurities has been intensively studied in the last 50 years\cite{WEB1980,ASO1987,DAV1987,DAVIES1989,TER1990,AWA1990,CAP1994,WAH2000,MIR2002,EST2003,pichler2004,CAR2011,REC2012,STE2013,CHR2015}.
In this framework, particular attention has been devoted to isovalent carbon-related defects, so-called G-centers \cite{YUK1965,SPR1968,BEA1970,JON1973,YUK1973,NOO1976,KIR1976,FOY1981,DAV1981,DAV1988,SAU1983,MAG1984,WEB1986,AWA1990,BEN1991,DAV1991,KWO1995,LAV1999,HAY2004,DAV2006,CLO2006,LON2013}  (sometimes labeled as A-centers\cite{CLO2005,CLO2006}) originally highlighted in carbon-rich Si samples undergoing high-energy irradiation with electrons, protons, neutrons, and gamma rays, followed by high temperature annealing. Although the intimate composition of this light emitter has been questioned for a long time \cite{THON1981,THON1984,DAV1981b,SONG1990,LEA1997}, it is now commonly accepted that it originates from a substitutional-interstitial carbon pair (C$_{S}$-C$_{I}$) interacting with an interstitial silicon (Si$_{I}$) \cite{CAP1998,MAT2002,POT2006,WAN2014,WAN2014b,TIM2017}.

The relevance of the G-center for optoelectronics has been highlighted in many works and it relies on some strategic aspects of its opto-electronic properties: (i) emission at 969 meV ($\sim$1280 nm) with a limited broadening of few meV, matching the important optical telecommunications wavelength O-band spreading between 1260-1360 nm; (ii) electrical injection of carriers, allowing for electro-luminescent devices \cite{CAN1987,CAN1989,ROT2007,MUR2011}; (iii) stimulated emission \cite{CLO2005,CLO2006}; (iv) high temperature emission (above liquid nitrogen) and eventually at room-temperature \cite{CLO2006b}; (v) ease of fabrication of high densities of G-centers via implantation of carbon ions, annealing and irradiation \cite{ROT2007,BER2012,BER2012b,BER2015}. It is worth stressing that differently from conventional III-V homo-epitaxial substrates, available only in small sizes (a few inches at most), or diamond substrates (a few millimeters), silicon-wafers can be as large as 12 inches and by far less expensive. Moreover, for silicon-based optical and electronic devices implementation, 12 inches silicon-on-insulator substrates are also available in a wide range of specifications of thickness and doping. 

In spite of the relevance of this topic, there are still many points to be clarified concerning the carrier dynamics in G-centers. For instance, the basic question of the lifetime remains unanswered, albeit of crucial importance for the brightness of single photon sources based on G-centers. For that purpose, we performed detailed photoluminescence (PL) experiments as a function of excitation energy, incident power, irradiation fluence and temperature in order to study the impact of radiative and non-radiative recombination channels on the spectrum, yield and lifetime of the G-center.

The paper is organized as follows: in Part \ref{sec:EXP}, we describe the sample fabrication (section \ref{sec:EXP_Sample}) and the experimental setups used for optical spectroscopy (section \ref{sec:EXP_setup}). Part \ref{sec:RES} is devoted to our results and their interpretation: we present the characterization of the G-centers sample (section \ref{sec:PLE}), PL measurements as a function of incident power revealing the saturation of the G-centers emission (section \ref{sec:FILLING}), time-resolved PL experiments unraveling the 6 ns-lifetime at low temperature (section \ref{sec:TIMERES}), the analysis of the phonon sideband leading to an estimation of 1.6$\pm$0.1 $\angstrom$ for the spatial extension of the electronic wave function in a G-center (section \ref{sec:SIDEBAND}), and finally a temperature-dependent study for extracting the radiative lifetime as a function of temperature (section \ref{sec:TEMPERATURE}). Part \ref{sec:conclu} is the general conclusion.

\section{SAMPLE DESCRIPTION AND EXPERIMENTAL SETUP}\label{sec:EXP}

\subsection{Sample fabrication}\label{sec:EXP_Sample}

Following a well-established procedure \cite{BER2012}, we implanted a 220 nm thick silicon-on-insulator wafer with a fluence of 2$\cdot$10$^{14}$ cm$^{-2}$ carbon ions (the beam energy was 36 keV, resulting in a 100 nm of the projected carbon range). The sample was annealed in N$_{2}$ atmosphere for 20 sec at 1000$^{\circ}$ C for removing the radiation damage.

Five different areas of 25$\times$25 $\mu$m$^{2}$ size were then implanted with protons, using fluences of 0.1, 0.3, 1, 3 and 9$\cdot$10$^{14}$ cm$^{-2}$. The implantation energy was set to 2.25 MeV.

\subsection{Optical spectroscopy}\label{sec:EXP_setup}

In our experimental setup, the sample was held on the cold finger of a closed-cycle cryostat for temperature-dependent measurements from 10 K to room temperature. The optical illumination was provided either by a cw HeNe laser at 632 nm (1.96 eV), by a cw laser diode at 532 nm (2.33 eV), or by a pulsed laser diode emitting at 532 nm with a repetition rate of 20 MHz. The excitation laser was focused onto the sample with a microscope objective (NA = 0.75), after reflection on a steering mirror for operating our scanning confocal microscope. The PL response was collected by the same objective.

For cw-detection, the PL signal was dispersed in a f=300 mm Czerny-Turner monochromator, equipped with a 600 grooves/mm grating blazed at 1600 nm, and recorded with an InGaAs array, with a quantum efficiency of 80$\%$ at 1300 nm, over integration times of 60 s.

For time-resolved measurements, the PL signal was detected by an InGaAs photodiode with a cut-off detection at 1700 nm, after spectral selection with a long-pass filter at 1250 nm. The additional use of 20 nm-bandpass filters allowed to spectrally address different parts of the emission spectrum of G-centers. The temporal decay was recorded by means of time-correlated-single-photon counting measurements with an overall temporal resolution of 400 ps.

For photoluminescence excitation (PLE) measurements, we used a cw-TiSa oscillator and a pseudo-cw source (super-continuum Fianium SC400-4) filtered by a holographic tunable bandpass filter (Photon etc) with a bandwidth of 2 nm. The average power density was monitored when tuning the excitation energy but remained on the order of 4-10 kW.cm$^{-2}$ all throughout the excitation window. The data are normalized to a constant power density of 10 kW.cm$^{-2}$. 

\section{RESULTS}\label{sec:RES}

In this section, we present our results that elucidate some of the fundamental opto-electronic properties which were not addressed by means of the experimental facilities available at early stages of the G-center investigations. We first describe the characterization of the G-centers sample (Section \ref{sec:PLE}), then power-dependent PL measurements showing the saturation of the G-centers emission (Section \ref{sec:FILLING}), followed by the first time-resolved PL measurements unraveling the decay dynamics on a 6-ns time-scale at low temperature (Section \ref{sec:TIMERES}), and demonstrating in the temporal domain the existence of phonon side-bands which are discussed in details in Section \ref{sec:SIDEBAND}, and eventually the temperature-dependent experiments (Section \ref{sec:TEMPERATURE}) focusing on the emission energy, zero-phonon line (ZPL) width, PL signal intensity, and recombination decay time.

\subsection{Characterization of the G-centers sample}\label{sec:PLE}
\begin{center}
\begin{figure*}
\includegraphics[width=0.9\textwidth]{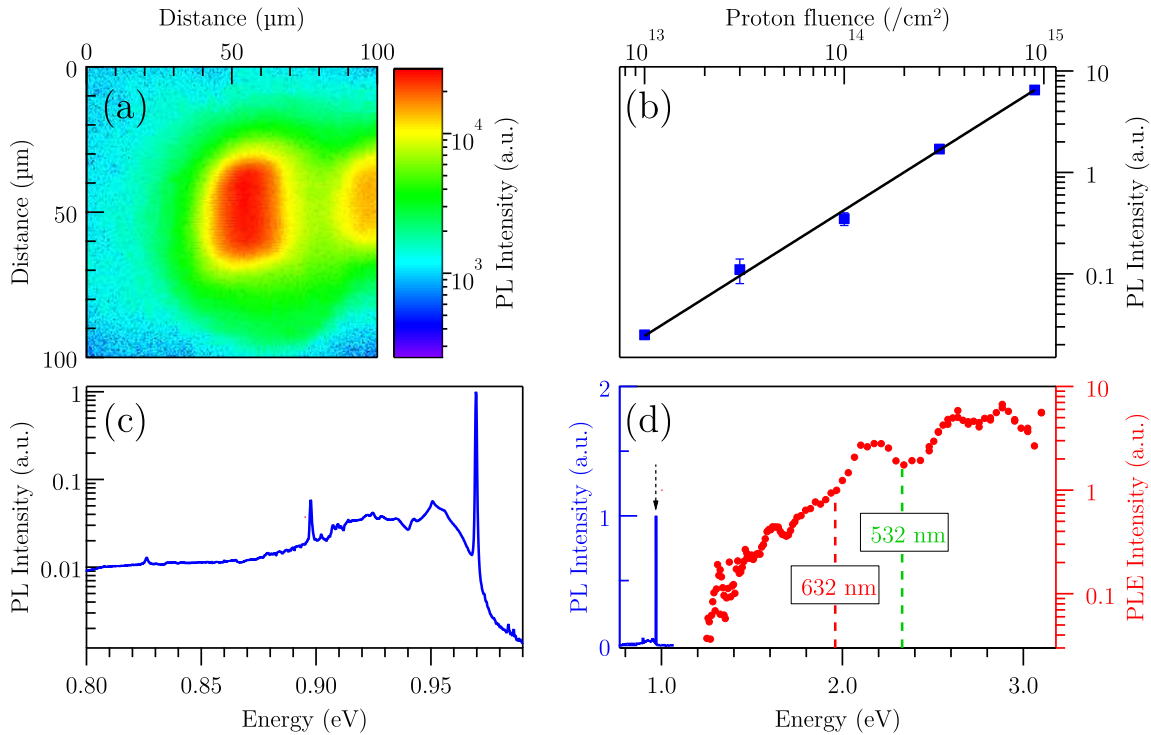}
\caption{(a) Photoluminescence (PL) raster scan of the pad irradiated with a proton fluence of 9$\cdot$10$^{14}$ cm$^{-2}$, for a cw-excitation energy of 2.33 eV, at 10 K. (b) PL signal intensity versus proton fluence. (c) PL signal intensity vs detection energy, on a semi-log scale, for a cw-excitation energy of 2.33 eV in the center of the pad seen in (a). (d) PL signal intensity (red symbols) vs excitation energy, for a central detection energy of 969 meV (black dashed arrow on the PL spectrum recalled in solid blue line) in the same ensemble of G-centers in silicon at 10 K, for a constant incident power of 10 kW.cm$^{-2}$.} 
\label{fig1}
\end{figure*}
\end{center}

As described in the literature, the best procedure for the generation of a high density of G-centers in silicon requires a two-step procedure: (i) introduction of carbon atoms in the silicon matrix, and (ii) irradiation in order to kick a carbon atom into an interstitial site, next to a substitutional carbon \cite{BER2012,BER2012b,BER2015}. As a matter of fact, starting from a given carbon concentration in a silicon sample, one expects the concentration of G-centers to increase with the irradiation fluence.

We first characterized the impact of the proton implantation by mapping the PL signal intensity in five areas implanted with different proton fluences. Fig.\ref{fig1}(a) displays the PL raster scan performed on the pad irradiated with a proton fluence of 9$\cdot$10$^{14}$ cm$^{-2}$ (part of the next pad being observable on the right side). The strong increase of the PL signal intensity in the implanted region demonstrates a dramatic influence of the irradiation. 

We further analyzed the PL signal intensity as a function of the proton fluence. The results are plotted in Fig.\ref{fig1}(b) on a log-log scale. The increase of the G-centers emission with the proton fluence appears slightly superlinear, with a power law of exponent 1.25$\pm$0.05. The G-centers concentration increases indeed with the proton fluence, with a nonlinear behavior possibly steming from the complex nature of the defect involving two carbon atoms, which was also reported in the literature in the regime of the low implanted densities \cite{BER2012}.

Fig.\ref{fig1}(c) displays the normalized PL signal intensity as a function of detection energy, for a cw-excitation energy of 2.33 eV in the center of the pad [Fig.\ref{fig1}(a)]. The sharp and most intense emission line centered at 969 meV stems from the ZPL of the G-center, i.e. the direct radiative recombination without phonon emission. At lower energy, we observe an additional component related to phonon-assisted recombination, as later discussed in Sections \ref{sec:TIMERES} and \ref{sec:SIDEBAND}.

Although some absorption measurements around the ZPL energy were reported in the literature \cite{DAVIES1989}, there is to the best of our knowledge no PLE spectroscopy of the G-centers. PLE provides a combined information on the two processes of absorption and carrier relaxation, and its knowledge is important for characterizing the spectral dependence of the effective pumping efficiency for a given incident power. Under excitation above the bandgap, one expects the PLE spectrum to largely reflect the silicon absorption.

Red symbols in Fig.\ref{fig1}(d) label the PLE spectrum in an ensemble of G-centers in silicon at 10 K. It was recorded by monitoring the emission intensity around the ZPL (black dashed arrow in Fig.\ref{fig1}(d)), as a function of the excitation energy for a constant incident power of 10 kW.cm$^{-2}$ (i.e. in the linear regime, as detailed in section \ref{sec:FILLING}).

In the semi-log plot of the PLE spectrum in Fig.\ref{fig1}(d), we first observe that the PL signal increases by two orders of magnitude when tuning the excitation energy from 1.2 to 3 eV. The fundamental bandgap of silicon is indirect, and at low temperature, it lies at an energy of about 1.16 eV. Consequently, our investigated range corresponds to a non-resonant, above bandgap excitation. In this case, the relaxation of the photo-generated carriers first consists in a non-radiative relaxation down to the extrema of the conduction and valence bands, followed by the capture in the G-centers. The excitation energy being always larger than the silicon bandgap in Fig.\ref{fig1}(d), the PLE spectrum essentially reproduces the absorption of silicon thin films \cite{wang_interpretation_1993}. 

\subsection{Saturation}\label{sec:FILLING}

In this section, we study the existence of saturation effects in the emission of G-centers. We demonstrate a sub-linear increase of the emission as a function of excitation power, and we estimate the saturation power from the quantitative interpretation of our data in an ensemble of G-centers.

\subsubsection{Sublinear power-dependence}\label{sec:FILLING_exp}
\begin{center}
\begin{figure}[h]
\includegraphics[width=0.46\textwidth]{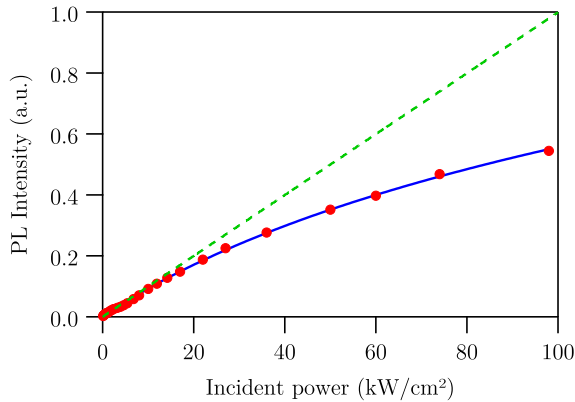}
\caption{Photoluminescence (PL) signal intensity of an ensemble of G-centers versus incident power, at 10 K. Experimental data (red circles) and fit (blue line) according to Eq.\eqref{eq1bis}, modelling the saturation of identical two-level systems excited by a Gaussian excitation spot, and assuming a saturation power $P_{\textrm{sat}}$=35 kW.cm$^{-2}$. The green dashed line shows a linear dependance as a function of incident power.} 
\label{fig2}
\end{figure}
\end{center}

We measured the dependence of the PL signal intensity of an ensemble of G-centers, as a function of the incident power $P$, at 10K, for a cw-excitation energy of 1.96 eV. The experimental data are plotted as red circles in Fig.\ref{fig2}. At low incident power ($P<$ 20 kW.cm$^{-2}$), the emission intensity increases quasi-linearly with $P$. In contrast, for $P\gtrsim$ 20 kW.cm$^{-2}$, a sublinearity of the PL signal intensity is clearly resolved.

We checked that there was no thermal effect biasing our power-dependent experiments under strong excitation, since a temperature rise induces a decrease of the PL signal intensity together with a thermal shift and broadening, as will be discussed later in Section \ref{sec:TEMPERATURE}. Thermal shift and broadening being absent in our power-dependent experiments, we conclude that the sublinearity of the emission intensity in Fig.\ref{fig2} is the signature for saturation effects in G-centers, that we analyze quantitatively below.

\subsubsection{Saturation of an ensemble of two-level systems}\label{sec:FILLING_th}

In the following, we assume that, in our experiments, an ensemble of G-centers is excited by a laser spot with a Gaussian profile, that all G-centers are identical, and that their emission intensity follows a standard saturation curve with a saturation power density $P_{\textrm{sat}}$. Such a framework only provides a first approximation for the interpretation of our measurements, since it does not take into account, for instance, the different defect orientations in the sample, however it allows us to reach a first order estimate of $P_{\textrm{sat}}$.

When raising the incident power $P$, the G-centers at the spot center are the first to saturate. Nonetheless, the region comprising saturated G-centers becomes progressively larger when increasing $P$. However, at the spot periphery, there are always non-saturated G-centers. As a consequence, power-dependent measurements in an ensemble of G-centers cannot display the standard saturation curve expected for a single two-level system.

In order to be more quantitative, we calculated the emission intensity assuming a two-dimensional distribution of G-centers, which is a reasonable assumption given the 220 nm thickness of our silicon-on-insulator sample. The recorded PL signal intensity thus reads:
\begin{equation}
I_{\textrm{PL}}\propto\int_{0}^{\infty} 2\pi rdr \left(P_0e^{-r^2/w^2}\right)\frac{1}{1+\frac{P_0e^{-r^2/w^2}}{P_{\textrm{sat}}}}
\label{eq1}
\end{equation}

where $P_0$ is the power density at the center of our Gaussian laser spot of waist $w$, and the right term of the integrand the saturation function of a two-level system. $P$ being the average incident power over the laser spot area, one finds $P=P_0/ln2$. A straightforward integration results in:
\begin{equation}
I_{\textrm{PL}}=I_0 \ln\left(1+\frac{P}{(P_{\textrm{sat}}/ln2)}\right)
\label{eq1bis}
\end{equation}

with $I_0$ the emission intensity for an incident power of $P_{\textrm{sat}}(e-1)/ln2$.

In Fig.\ref{fig2}, a fit according to Eq.\eqref{eq1bis} provides an excellent agreement with our experimental data, by taking for our fitting parameter $P_{\textrm{sat}}$ a value of 35$\pm$7 kW.cm$^{-2}$.

The quantitative interpretation of our power-dependent experiments shows that the saturation of the emission can be resolved by ensemble measurements in G-centers, thus leading to an estimation of the saturation power. Such a strategy is specific to point defects where the assumption of an identical saturation power $P_{\textrm{sat}}$ for all emitters is crucial. This hypothesis is not met in other nanostructures, such as epitaxial quantum dots or nanocrystals. In these latter cases, where inhomogeneous line broadening arising from size dispersion dominates, the fluctuations of the fundamental properties (lifetime, dephasing time) are important enough to prevent the observation of saturation effects by ensemble measurements.

The 35$\pm$7 kW.cm$^{-2}$ value of $P_{\textrm{sat}}$ can be used to roughly estimate the order of magnitude of the carrier capture volume in G-centers. Assuming a simple Poissonian model for the level occupation probability \cite{grundmann,ABB2009}, the average number of excitons within a G-center is one at saturation. Provided a lifetime of $\sim$6 ns as obtained by time-resolved experiments (see the following section) and an absorption length of $\sim$5$\cdot$10$^{-4}$ cm for a laser excitation at 1.96 eV, the steady-state carrier density results in $\sim$2$\cdot$10$^{17}$ cm$^{-3}$. Assuming a spherical geometry of the extrinsic center, the capture volume is represented by the inverse of this carrier density, and it leads to a capture radius of 20$\pm$2 nm.

This value is quite similar to the one found in extrinsic centers in III-V alloys \cite{DOT2015}. While point defects represent a modification of the crystal lattice at the atomic scale, the capture volume is strikingly much wider than the defect size, approximately two orders of magnitude larger than the extension of the electronic wave-function within the G-center (see section \ref{sec:SIDEBAND}). The capture radius further gives an interesting estimate of the effective volume where the captured charge carriers may influence the optical response via spectral diffusion \cite{abbarchi_spectral_2008,ABB2009}, this phenomenon providing an important contribution to the ZPL broadening, as later discussed in section \ref{sec:TEMPERATURE_zpl}.
\subsection{Recombination dynamics}\label{sec:TIMERES}

As a marker of the residual carbon concentration in silicon, G-centers were extensively studied decades ago, in the prospect of growing bulk silicon crystals as pure as possible \cite{DAVIES1989}. Surprisingly, the prominent question of the lifetime value remains unanswered, primarily because of the limited temporal resolution of the earlier experiments, so that only the upper bound of 4 $\mu$s is mentioned in the review of Davies \cite{DAVIES1989} (Thonke \textit{et al.} having nevertheless inferred to the upper bound of 10 ns in Ref.\onlinecite{THON1981}). In the following, we unravel the recombination dynamics by means of time-resolved PL measurements with a temporal resolution of 400 ps.

\subsubsection{Spectrally-selective time-resolved PL measurements}\label{sec:TIMERES_spec}
\begin{center}
\begin{figure}[h]
\includegraphics[width=0.49\textwidth]{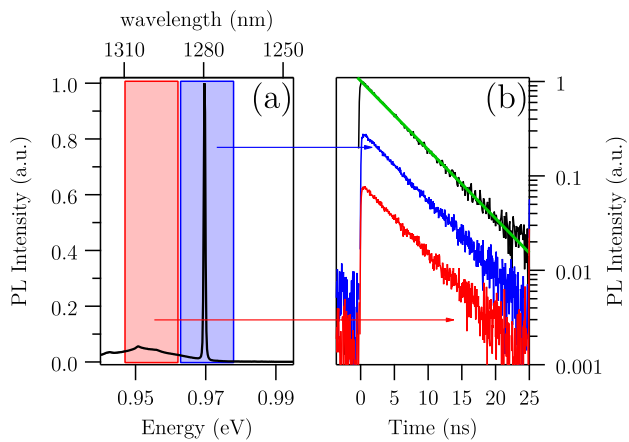}
\caption{(a) PL spectrum of G-centers in silicon at 10 K. The blue and red shaded areas indicate the spectral width of the two bandpass filters used for spectrally-selective time-resolved measurements of the ZPL and phonon sideband, respectively. (b) Time-resolved PL signal intensity for the whole spectrum (black line), the ZPL only (blue line), and the phonon sideband only (red line). The average incident power is 1 kW.cm$^{-2}$. The green line indicates an exponential decay with a time constant of 5.9 ns.}
\label{fig4}
\end{figure}
\end{center}

The black line in Fig.\ref{fig4}(b) is the time-resolved trace of the PL signal intensity, spectrally-integrated over the whole emission spectrum of G-centers, from 1250 to 1700 nm (see section \ref{sec:EXP_setup}). On the semi-log scale of Fig.\ref{fig4}(b), we observe that the decay of the PL signal is purely exponential over the two measured decades, with a characteristic time constant of 5.9 ns. This lifetime is slightly longer than the 1.3 ns-value in InAs quantum dots \cite{gerard,dousse}, but shorter than the 11 ns-one in the prototypical NV center in diamond \cite{gruber}. The isolation of single G-centers would thus open the prospect of obtaining bright single photon emitters in silicon.

An interesting and original insight into the optical response of G-centers is reached by performing spectrally-selective time-resolved PL experiments. By means of bandpass filters, we measured the recombination dynamics of the ZPL [blue shaded area in Fig.\ref{fig4}(a)], and of the low-energy sideband [red shaded area in Fig.\ref{fig4}(a)]. The corresponding time-resolved traces are plotted as blue and red lines in Fig.\ref{fig4}(b), respectively. They are strictly identical with the same time-constant of 5.9 ns found above. This observation indicates the common nature of these two recombination channels.

Although the low-energy part of the PL spectrum was early identified as coming from phonon-assisted recombination in analogy to the general phenomenology in point defects \cite{DAVIES1989}, spectrally-selective time-resolved PL measurements provide here a powerful way for establishing that recombination processes leading to photons of different energy share the same microscopic origin. In fact, the recombination dynamics of an electronic two-level system in a phonon bath occurs either via direct radiative recombination (corresponding to the ZPL), or via phonon-assisted recombination (corresponding to the phonon sidebands). Whatever the number of emitted or absorbed phonons, all these mechanisms contribute in parallel to the recombination dynamics of the excited state of the two-level system. As a matter of fact, the lifetime depends on the electronic dipole and on the strength of the electron-phonon interaction. Still, whatever the detuning with the ZPL, one expects the very same dynamics when performing time-resolved PL measurements. In other words, the decay time of the ZPL and phonon sidebands must be equal. This general property is surprisingly very poorly documented in the literature \cite{ABB2008,intervalley}. Fig.\ref{fig4} nicely illustrates it in the context of the optical response of G-centers in silicon.

Before further analyzing the spectrum of the phonon sideband in G-centers (see section \ref{sec:SIDEBAND}), we present below our measurements of the lifetime as a function of the proton fluence.

\subsubsection{Lifetime versus proton fluence}\label{sec:TIMERES_dose}
\begin{center}
\begin{figure}[h]
\includegraphics[width=0.46\textwidth]{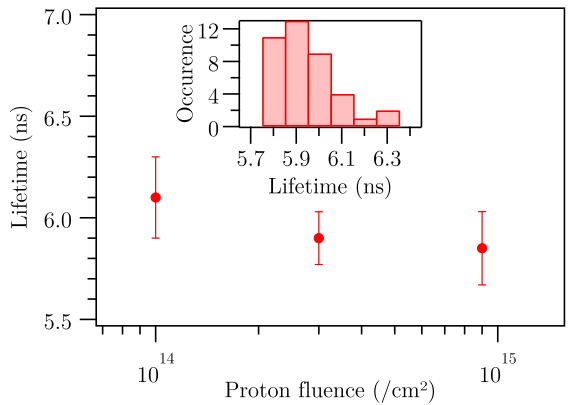}
\caption{Average lifetime (red circles) for the three highest proton fluences. Inset : lifetime histogram for the area implanted at $3\cdot$10$^{14}$ \textrm{proton}.cm$^{-2}$.}
\label{fig5}
\end{figure}
\end{center}

Although the absolute estimation of the areal density of G-centers is still currently very difficult, especially in the absence of single G-centers spectroscopy, we studied the possible influence of the G-centers concentration on the recombination lifetime. Our motivation was to examine if the close proximity of G-centers could induce any non-radiative relaxation channel.

In order to investigate this point, we performed time-resolved PL experiments on 40-50 different locations for each of the pad irradiated by a given proton fluence (see section \ref{sec:EXP_Sample}). Because of the limited incident power of our pulsed laser diode (average incident power of 1 kW.cm$^{-2}$), only the three highest proton irradiation fluences were accessible. The results are summarized in Fig.\ref{fig5}, where the inset shows the histogram of the measured lifetimes for the area implanted with $3\cdot$10$^{14}$ \textrm{proton}.cm$^{-2}$. The symbols in Fig.\ref{fig5} correspond to the mean value of the recorded statistical distributions, with the error bars representing the standard deviations. Although the mean lifetime values decrease from 6.1 to 5.9 ns on raising the proton dose, the variation is still within the experimental error bar of $\pm$0.2 ns. Consequently, no definite conclusion can be drawn on a possible influence of the G-centers concentration on their lifetime, within the 1-9$\cdot$10$^{14}$ \textrm{proton}.cm$^{-2}$ dose range.
\subsection{Phonon-assisted recombination}\label{sec:SIDEBAND}

The spectrally-selective time-resolved PL measurements brought a direct illustration, in the time domain, of the common microscopic origin of the different recombination paths highlighted as shaded areas in Fig.\ref{fig4}. Although the low-energy sideband was early interpreted as arising from phonon-assisted emission, we revisit below the emission spectrum of G-centers in the light of the modern theoretical approaches allowing a non-perturbative calculation of the acoustic phonon sidebands \cite{KRU2002,VUO2016}.

\subsubsection{Phonon sidebands}\label{sec:SIDEBAND_data}
\begin{center}
\begin{figure}[h]
\includegraphics[width=0.46\textwidth]{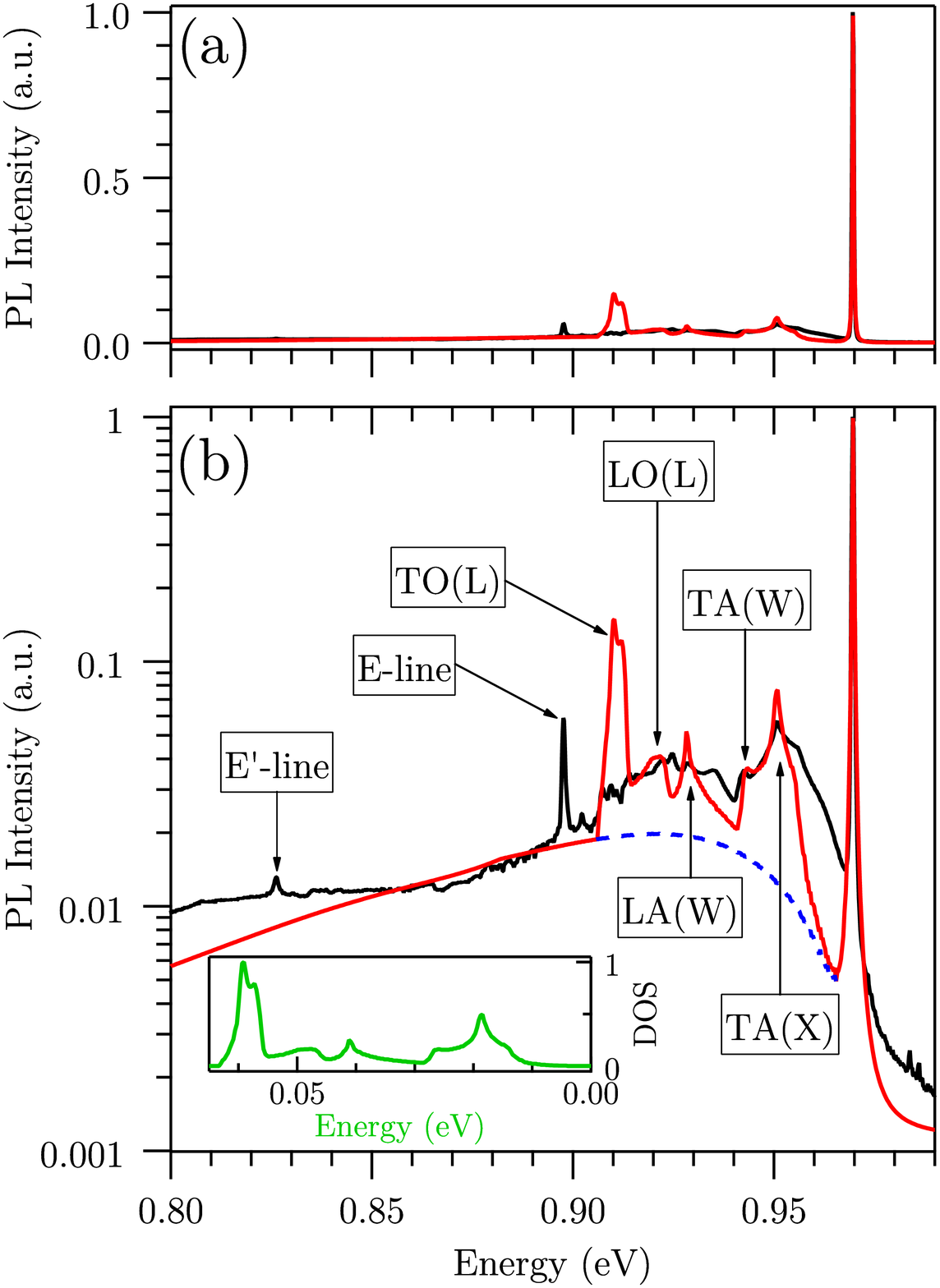}
\caption{PL spectrum of G-centers in silicon at 10 K for an excitation at 1.96 eV on a linear scale (a) and a semi-log scale (b). Experimental data (black line), calculated spectrum (red line). The blue dashed line is the calculation of the longitudinal acoustic phonon sideband with a point defect extension $\sigma$=1.6 $\angstrom$. Inset: phonon density of states in silicon versus energy.}
\label{fig3}
\end{figure}
\end{center}

The PL spectrum in an ensemble of G-centers at 10 K is plotted in Fig.\ref{fig3}, either on a linear scale [Fig.\ref{fig3}(a)] or on a semi-log scale [Fig.\ref{fig3}(b)], in a spectral domain covering a 0.2 eV-range around the 969 meV-energy of the ZPL. The black (red) line corresponds to the experimental data (calculated spectrum, respectively).

The PL spectrum is dominated by the narrow ZPL with a full width at half maximum (FWHM) of 0.3 meV, accompanied by a low-energy phonon sideband. While phonon emission gives rise to a red-shifted emission after phonon-assisted recombination, phonon absorption leads on the contrary to a blue-shifted emission with respect to the ZPL \cite{KRU2002,VUO2016}. At low temperature, the probability of phonon absorption is negligible compared to phonon emission, leading to the asymmetric emission spectrum at 10 K displayed in Fig.\ref{fig3}.

We now comment on the structure of the phonon sideband extending below 969 meV. We observe two structured broad peaks at 0.95 and 0.93 eV, followed by a sharp line at 0.89 eV, the so-called E line \cite{THON1981}, and a second one at 0.825 eV, the so-called E' line \cite{THON1981}.

The higher the phonon energy, the larger the detuning of the phonon replica with the ZPL. The relative maxima observed at 0.95 and 0.93 eV can thus be directly related to extrema of the phonon density of states implying specific phonon modes, as indicated in Fig.\ref{fig3}(b) and further discussed below. Very importantly, all these structures are superimposed to a broad pedestal arising from the recombination assisted by acoustic phonons \cite{KRU2002,VUO2016}. As a matter of fact, the quantitative interpretation of the acoustic phonon sideband allows a direct estimation of the spatial extension of the electronic wavefunction in the defect \cite{VUO2016}. We describe below the theoretical calculations implemented in the case of G-centers in silicon.

\subsubsection{Theoretical modelling}\label{sec:SIDEBAND_th}

In the framework of the theoretical approach derived from the Huang-Rhys model and developed for calculating the coherent nonlinear response in semiconductor quantum dots and carbon nanotubes \cite{KRU2002,vialla2014}, we have computed the sideband arising from the coupling to acoustic phonons in a defect inside a silicon matrix. Close to the zone-center (i.e. for small wavevectors), the deformation potential interaction is allowed only for LA phonons, while piezoelectric coupling is allowed for both LA and TA phonons \cite{KRU2002}. As silicon is a centro-symmetric material and thus non-piezoelectric, the only remaining coupling is the deformation potential for LA phonons \cite{cardona}. The emission spectrum is thus obtained by taking the Fourier transform of the time-dependent linear susceptibility $\chi (t)$ given by \cite{KRU2002}:
\begin{equation}
\chi (t)=\exp\left[\sum_{\textbf{k}}\left|\gamma_{\textbf{k}}\right|^2\left(e^{-i\omega(\textbf{k})t}-n(\textbf{k})\left|e^{-i\omega(\textbf{k})t}-1\right|^2-1\right)\right]
\label{eqchi}
\end{equation} 
where $\omega(\textbf{k})$ is the energy of a LA phonon of wavevector $\textbf{k}$, $n(\textbf{k})$ the corresponding Bose-Einstein phonon occupation factor. The dimensionless coupling strength $\gamma_{\mathbf{k}}$ reads: 
\begin{equation}
\gamma_{\mathbf{k}}=\frac{g^e_{\mathbf{k}}-g^h_{\mathbf{k}}}{\omega(\mathbf{k})}
\label{eqgamma}
\end{equation}
where $g^\alpha_{\mathbf{k}}$ is the coupling strength for electrons ($\alpha$=e) and holes ($\alpha$=h) given by:
\begin{equation}
g^\alpha_{\mathbf{k}}=G^\alpha_{\mathbf{k}}F^\alpha_{\mathbf{k}}
\label{eqg}
\end{equation}

$G^\alpha_{\mathbf{k}}$ is related to the electron-phonon interaction, and $F^\alpha_{\mathbf{k}}$ to the electronic wavefunction in the reciprocal space. More precisely, $F^\alpha_{\mathbf{k}}$ is the Fourier transform of the square modulus of the electronic wavefunction given by:
\begin{equation}
F^\alpha_{\mathbf{k}}=\int d^3\mathbf{r}|\Psi^\alpha(\mathbf{r})|^2e^{i\mathbf{k}.\mathbf{r}}
\label{eqF}
\end{equation}
where $\Psi^\alpha(\mathbf{r})$ is the wavefunction in the point defect. In order to obtain the typical extension of the electronic wavefunction in a G-center, we have taken a Gaussian of extension $\sigma$, which is assumed identical for both electron and hole, resulting in $F^\alpha_{\mathbf{k}}$=$\exp(-k^2\sigma^2/4)$.

As far as the electron-phonon interaction is concerned, only the deformation potential is relevant in our case since silicon is not piezoelectric, so that $G^\alpha_{\mathbf{k}}$ reads \cite{KRU2002}:
\begin{equation}
G^\alpha_{\mathbf{k}}=\frac{kD^\alpha}{\sqrt{2\varrho\hbar\omega(\mathbf{k})V}}
\label{eqG}
\end{equation}
with $D^\alpha$ the deformation potential, $\varrho$ the silicon density, and $V$ a normalization volume.

Since the present model is limited to linear terms in the electron-phonon interaction, the phonon-assisted broadening of the ZPL is not accounted for in our calculations \cite{KRU2002}, and the finite broadening of the ZPL has to be introduced phenomenologically by convoluting the emission spectrum with a Lorentzian line of FWHM $\Gamma_{ZPL}$. In the temporal domain, where the time-dependent susceptibility has the analytical expression of Eq.\eqref{eqchi}, this means multiplying $\chi(t)$ by an exponential function of time constant $2\hbar/\Gamma_{ZPL}$:
\begin{equation}
\widetilde{\chi(t)}=\chi(t)e^{-\Gamma_{ZPL}t/2\hbar}
\label{eqchibis}
\end{equation}

The calculated emission spectrum displayed in dashed blue line in Fig.\ref{fig3} is thus the Fourier transform of $\widetilde{\chi(t)}$ (the solid red line and the dashed blue one coincide when the latter is not visible).

In our calculations, we take for the deformation potential values $D^e$=10 eV and $D^h$=5 eV \cite{buin}, and only two parameters are free: the extension $\sigma$ of the electronic wave-function in the G-center, and the FWHM of the ZPL $\Gamma_{ZPL}$. $\Gamma_{ZPL}$ is a phenomenological broadening introduced in the model, since the latter does not account for the thermally-assisted broadening of the ZPL. Its value is adjusted in order to reproduce the ZPL, and in Fig.~\ref{fig3}, $\Gamma_{ZPL}$=0.3 meV (note that this value is the zero-temperature limit in Fig.\ref{fig7}(b)). However, it is obvious from Fig.\ref{fig3} that the sideband due to the longitudinal acoustic phonons (blue dashed line) does not bring by itself a full quantitative interpretation of the PL spectrum, but only a baseline on top of which appear the E and E' lines and the two bands centered at 0.95 and 0.93 eV.

In our modelling of the phonon-assisted recombination, one thus needs to go beyond longitudinal acoustic phonons in order to reach an estimation of the G-center extension $\sigma$. For that purpose, we added a contribution proportional to the phonon density of states in silicon [inset, Fig.\ref{fig3}(b)]. Such a procedure is a very crude attempt to take into account phonon modes other than longitudinal acoustic phonons, since it assumes a constant electron-phonon matrix element, irrespective of the exact form of the electron-phonon coupling and of the interaction selection rules. Moreover, it further assumes that phonon-assisted recombination is dominated by emission processes involving only one phonon. This hypothesis is more likely to be fulfilled at low temperature \cite{VUO2016}, as it is the case here. As a matter of fact, by adding a contribution proportional to the phonon density of states in silicon \cite{tomas2014}, we significantly improve the fit of our data [solid red line in Fig.\ref{fig3}]. By varying the G-center extension $\sigma$ and the weight of this additional contribution, we reach a fair agreement with $\sigma$=1.6$\pm$0.1 $\angstrom$. This number is smaller than the 2.3 $\angstrom$ spacing between nearest neighbors in silicon, and it is very close to the C-Si bound length in G-centers \cite{CAP1994,CAP1998,WAN2014,WAN2014b}. Our theoretical approach therefore provides an original method for estimating the spatial extension of the electronic wave function in a G-center.

From this analysis of the phonon sidebands, we eventually identify the different features observed in the PL spectrum. Namely, the two peaks at 0.95 and 0.94 eV correspond to extrema of the density of states related to transverse acoustic (TA) phonons at the X and W points of the Brillouin zone, respectively (see for instance Ref.\onlinecite{tomas2014} for the bandstructure and density of states of phonons in silicon). For the band centered at 0.93 eV, there is an overall agreement with the density of states, but not as precise as in the previous case. On the one hand, one perceives the longitudinal acoustic (LA) and optical (LO) phonons at the W and L points, respectively. On the contrary, the sharp maximum of the density of states due to transverse optical (TO) at the L point is completely missing in the experimental spectrum, suggesting a suppression of the corresponding phonon replica because of selection rules.

Conversely, the emission spectrum comprises a low-energy component extending from 0.97 to 0.95 eV, that is not reproduced by our theoretical fit. In analogy to the vibronic spectrum in NV centers \cite{alkauskas}, we tentatively attribute it to a defect-induced vibrational resonance, that does not come from the bulk phonons in silicon but from the atomic vibrations in the G-center itself.

As far as the E-line is concerned, the detuning with the ZPL is 72 meV, i.e. a value larger than the maximum of the phonon band-structure. It originates from a localized phonon mode because of the presence of the G-center, as expected for a defect lighter than the atoms of the crystal \cite{bjork}.

Both cases can not be reproduced by our calculations which solely rely on the phonon Bloch modes in a perfect silicon matrix. We believe that the extension of \textit{ab initio} theoretical treatments (such as in Ref.\onlinecite{alkauskas}) in G-centers should complete the picture of phonon-assisted recombination in G-centers. We nevertheless highlight that our original approach based on non-perturbative calculations of the acoustic phonon sideband and an \textit{ad hoc} inclusion of zone-edge phonons provide a direct estimation of the spatial extension of the electronic wave function in a G-center.
\subsection{Temperature-dependent photoluminescence spectroscopy}\label{sec:TEMPERATURE}

In the last part of the paper, we present the temperature-dependent PL measurements performed in G-centers, with a special emphasis on time-resolved experiments (section \ref{sec:TEMPERATURE_decay}), followed by the analysis of the thermal red-shift of the ZPL (section \ref{sec:TEMPERATURE_shift}), its temperature-broadening (section \ref{sec:TEMPERATURE_zpl}), before the comparison of the temperature dependence of the PL signal intensity (section \ref{sec:TEMPERATURE_int}) and recombination lifetime (section \ref{sec:TEMPERATURE_decaytime}).
\subsubsection{Temperature-dependent recombination dynamics}\label{sec:TEMPERATURE_decay}
\begin{center}
\begin{figure}[ht]
\includegraphics[width=0.5\textwidth]{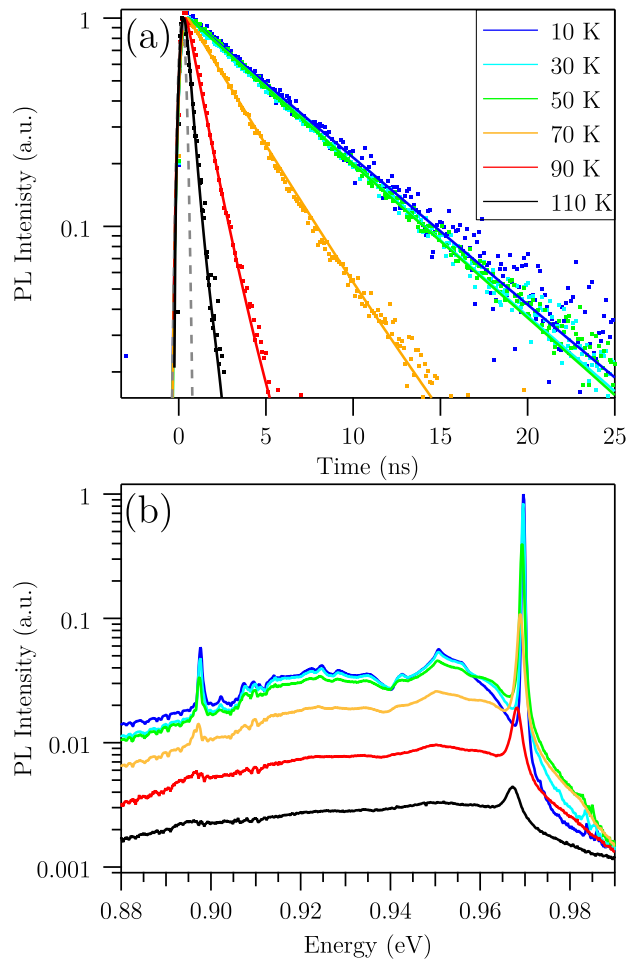}
\caption{Temperature-dependent PL spectroscopy in an ensemble of G-centers in silicon, in the temperature range 10-110 K. (a) Time-resolved experiments: data (symbols), and fit (solid line) after convolution of the system response function (dashed line) with an exponential decay. (b) Emission spectra.}
\label{fig6}
\end{figure}
\end{center}

Let us start with the recombination dynamics as a function of temperature. Fig.\ref{fig6}(a) displays the time-resolved PL measurements on a semi-log scale, for temperatures ranging from 10 to 110 K. Below 50 K, the PL decay is almost unchanged with identical temporal traces. In contrast, from 70 to 110 K, the recombination strongly fastens so that the estimation of the lifetime requires to take into account the system response function. For the sake of consistancy, we systematically convoluted the system response function [dashed line in Fig.\ref{fig6}(a)] with an exponential decay for adjusting our temperature-dependent data. The lifetime decreases from 5.9 ns below 50 K, to 0.5 ns at 110 K, with intermediate values of 3.2 and 1.1 ns at 70 and 90 K, respectively.

Generally speaking, the PL decay time gets shorter on raising the temperature because of the thermally-assisted decrease of either the radiative lifetime or the non-radiative one. The temperature dependence of the radiative recombination time was identified as an intrinsic feature in semiconductor materials having a translational invariance along at least one direction, namely bulks, quantum wells, and quantum wires or carbon nanotubes \cite{andreani,rosales,berger2007}. In zero-dimensional nanostructures such as epitaxial quantum dots and colloidal nanocrystals, the radiative lifetime no longer varies with temperature because of the suppression of thermalization effects along the excitonic dispersion. The same phenomenology is expected in point defects, suggesting that the faster recombination dynamics at high temperature in G-centers only comes from thermally-assisted non-radiative recombination [Fig.\ref{fig6}(a)].

In order to check this important point, and following a well-established method \cite{andreani,rosales}, we performed complementary measurements of the absolute PL signal intensity as a function of temperature [Fig.\ref{fig6}(b)]. On the semi-log scale of Fig.\ref{fig6}(b), one sees that the ZPL red-shifts and broadens on raising the temperature, with a global reduction of the PL signal intensity by approximately two decades from 10 to 110 K. Moreover, the sharp features of the phonon sideband gradually disappear as a result of the ZPL broadening \cite{VUO2016}. Finally, the increasing probability of phonon absorption gives rise to PL emission at higher energy than the ZPL, so that the asymmetry of the PL spectrum is smoothly reduced at high temperatures.

In the following, we perform the quantitative analysis of the whole set of data displayed in Fig.\ref{fig6}.
\begin{center}
\begin{figure}[h]
\includegraphics[width=0.425\textwidth]{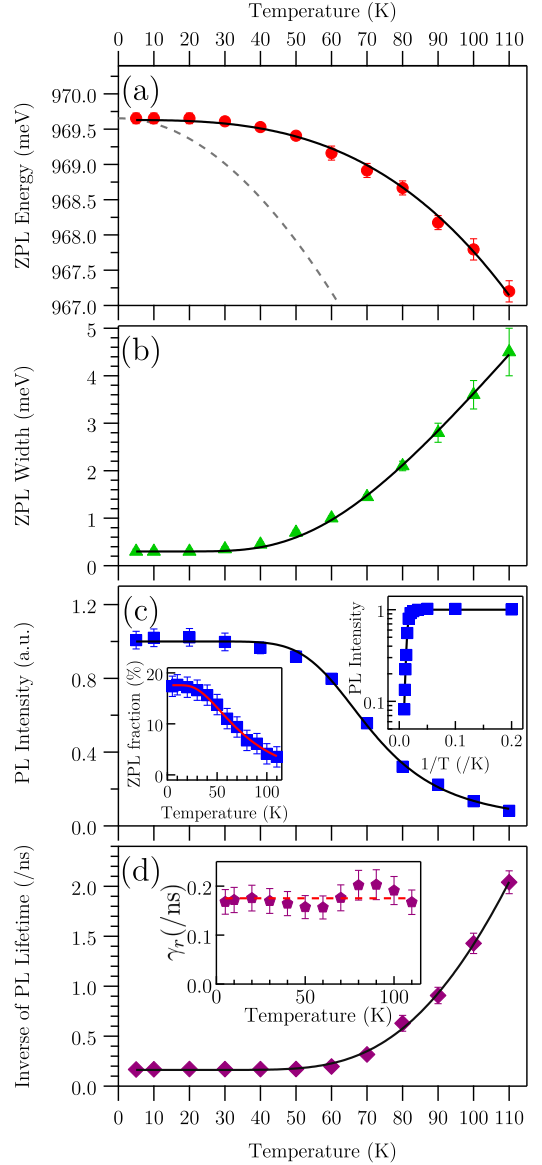}
\caption{(a) Temperature-dependence of the ZPL energy: data (symbols), Varshni's law for bulk silicon (dashed line), and cubic fit according to Eq.\eqref{eq7a} (solid line). (b) Broadening of the ZPL: data (symbols), fit according to Eq.\eqref{eq7b} (solid line). (c) PL signal intensity: data (symbols), fit according to the Arrhenius law of Eq.\eqref{eq7c} (solid line). Right inset: same graph plotted as a function of 1/T. Left inset: ZPL fraction as a function of temperature. (d) Inverse of the PL decay time: data (symbols), fit according to Eq.\eqref{eq7d} (solid line). Inset: radiative rate estimated from the product of the PL signal intensity [Fig.\ref{fig7}(c)] times the recombination rate [Fig.\ref{fig7}(d)]. The dashed horizontal line is a guide for the eye.}
\label{fig7}
\end{figure}
\end{center}
\subsubsection{Thermal red-shift}\label{sec:TEMPERATURE_shift}

The energy of the ZPL is plotted as a function of temperature in Fig.\ref{fig7}(a). We first compare the measured variations with the bandgap shift of bulk silicon [dashed line in Fig.\ref{fig7}(a)] \cite{alex1996}. The obvious disagreement accounts for a modification of the electron-phonon interaction in G-centers compared to the silicon matrix.

In bulk silicon, the bandgap variations originate from (i) the lattice expansion with temperature, leading to the linear decrease of the bandgap energy at high temperature, and from (ii) a 90 meV-renormalization of the bandgap energy at low temperature because of zero-point fluctuations \cite{cardona_isotope_2005}. The net result gives a bended curve, which may be described by various phenomenological expressions, the most common one being the Varshni's law \cite{cardona_isotope_2005}.

The distinct temperature dependence of the ZPL energy suggests modified zero-point fluctuations in G-centers, which understanding is beyond the scope of this work since it requires detailed calculations of the electron-phonon coupling in this type of point defects. Eventually, we note that the thermal red-shift of G-centers is fairly reproduced [solid line in Fig.\ref{fig7}(a)] by the following polynomial expression:
\begin{equation}
E_{\textrm{ZPL}}=E_0-AT^p
\label{eq7a}
\end{equation}

with $E_0$=969.6$\pm$0.1 meV, $A$=1.9$\pm$0.2$\cdot$10$^{-6}$meV.K$^{-3}$, and $p$=3$\pm$0.1. The latter value is consistent with the 2-3.3 range measured for $p$ in tens of semiconductors \cite{cardona_isotope_2005}.
\subsubsection{Broadening of the Zero-Phonon Line}\label{sec:TEMPERATURE_zpl}

The thermal red-shift of the ZPL comes along with a pronounced broadening on raising the temperature. As displayed in Fig.\ref{fig7}(b), the ZPL width at low temperature is about 0.3 meV, and it stays rather constant until 20 K. At larger temperature (about 30 K) we observe an increasing trend of the ZPL width reaching the value of 4.5 meV at 110 K.

The ZPL broadening is well fitted [solid line in Fig.\ref{fig7}(b)] with the expression \cite{rudin}: 
\begin{equation}
\Gamma=\Gamma_0+ae^{-\Omega/k_BT}
\label{eq7b}
\end{equation}

with $\Gamma_0$=0.3 $\pm$0.05 meV, $a$=34$\pm$5 meV, and $\Omega$=21$\pm$2 meV.

$\Gamma_0$ is the zero-temperature limit of the ZPL width. Since we performed ensemble measurements, the ZPL is probably inhomogeneously broadened so that the 0.3 meV-value of $\Gamma_0$ only brings an upper bound for the homogeneous linewidth. With a lifetime of 5.9 ns at 10 K, the radiative broadening is in the sub-$\mu$eV range, possibly suggesting the presence of spectral diffusion in addition to the inhomogeneous broadening due to ensemble measurements \cite{berthelot}.

The second term stems from phonon-assisted broadening. The exponential increase is reminiscent of the Bose-Einstein occupation factor of phonons $n(T)$ in the low-temperature regime ($\Omega\ll k_BT$) since the probability of phonon absorption is proportional to $n(T)$. Although Eq.\eqref{eq7b} is formally close to the usual expressions used in bulks, quantum wells or quantum wires \cite{rudin}, phonon dephasing in a zero-dimensional system, such as a point defect, can not be described within the same framework. As a matter of fact, the thermally-assisted broadening in epitaxial quantum dots was interpreted as an activation of the fluctuating environment responsible for spectral diffusion \cite{favero}. We note that the 21 meV-value for the mean phonon energy $\Omega$ concurs with the maximum of the phonon density of states around 20 meV [Fig.\ref{fig3}(b), inset], corresponding to the TA(X) mode, thus indicating the predominance of this phonon mode in the ZPL broadening.

Complementary measurements of the homogeneous broadening in single G-centers will be required in order to further elucidate the mechanisms controlling the ZPL width, and in particular the impact of spectral diffusion.
\subsubsection{Temperature dependence of the PL signal intensity}\label{sec:TEMPERATURE_int}

We now discuss the temperature dependence of the PL signal intensity integrated from 0.82 to 1 eV. As already commented above, the emission intensity decreases by more than one order of magnitude from 10 to 110 K. A fair agreement is reached [solid line in Fig.\ref{fig7}(c)] by means of an Arrhenius fit:
\begin{equation}
I(T)=\frac{I(0)}{1+Be^{-\frac{E_{a}^{(1)}}{k_BT}}}
\label{eq7c}
\end{equation}

with $B$=700$\pm$200, and $E_{a}^{(1)}$=41$\pm$5 meV. The right inset in Fig.\ref{fig7}(c) displays the same graph plotted as a function of 1/T where we better observe that a single activation energy well accounts for the experimental data. The 41 meV-value for the activation energy $E_{a}^{(1)}$ is consistent with the literature \cite{who}. We note that it strongly deviates from the confinement energy in comparison to other nanostructures, such as epitaxial quantum dots \cite{leru}. It will be further discussed in the light of the temperature dependence of the lifetime.

We also evaluated the fraction of the PL signal intensity emitted in the ZPL, also called Debye-Waller factor $\theta(T)$. The results are plotted in the left inset of Fig.\ref{fig7}(c). The ZPL fraction decreases from 18\% at 10 K to a few percents at 110 K. Assuming that the whole phonon bath can be approximated by a single phonon of energy $\Lambda$, the temperature dependence of the Debye-Waller factor is given by \cite{saikan}:
\begin{equation}
\theta(T)=\exp\left(-\xi^2\coth\left(\Lambda/2k_BT\right)\right)
\label{eq7cbis}
\end{equation}
where $\xi$ is the coupling strength of the linear electron-phonon interaction. In the left inset of Fig.\ref{fig7}(c), $\theta(T)$ is adjusted by taking $\xi$=1.3$\pm$0.05 and $\Lambda$=11$\pm$2 meV. The latter value is smaller than the 21 meV-mean phonon energy $\Omega$ entering Eq.\eqref{eq7b}, indicating a different origin of the ZPL broadening and thermal decrease of the Debye-Waller factor. While the thermal increase of the ZPL width is mostly determined by the TA(X) mode (see section \ref{sec:TEMPERATURE_zpl}), the decrease of the ZPL fraction may be due to the defect-induced vibrational resonance \cite{alkauskas}, discussed above as probably responsible for the low-energy component extending from 0.97 to 0.95 eV.

As far as the coupling strength $\xi$ is concerned, it directly determines the ZPL fraction at zero temperature, since $\theta(0)=\exp(-\xi^2)$. The coupling strength increases either with the electronic confinement or with the electron-phonon interaction so that it is complicated to compare point defects with quantum dots, or point defects in other materials. Still, given the fact that G-centers are shallow levels compared to the silicon bandgap, the coupling strength $\xi$ in G-centers is rather large comparing to deep levels in hexagonal boron nitride ($\xi$=1.1) \cite{VUO2016}, or NV centers in diamond ($\xi$=1.87) \cite{alkauskas}.
\subsubsection{Temperature dependence of the PL decay time}\label{sec:TEMPERATURE_decaytime}

We finally analyze the PL decay time $\tau$ as a function of temperature. The inverse of the lifetime is displayed in Fig.\ref{fig7}(d), and our data are fairly reproduced by the following expression:
\begin{equation}
\frac{1}{\tau}=\frac{1}{\tau_0}+Ce^{-\frac{E_{a}^{(2)}}{k_BT}}
\label{eq7d}
\end{equation}

with $\tau_0$=5.9$\pm$0.1 ns, $C$=120$\pm$20 ns$^{-1}$, and $E_{a}^{(2)}$=39$\pm$5 meV.

The activation energy $E_{a}^{(2)}$ is identical to $E_{a}^{(1)}$ (from Eq.\eqref{eq7c}) within our experimental error. This is an important point in the prospect of extracting the radiative lifetime as a function of temperature.

On the one hand, the time-integrated PL signal intensity [Fig.\ref{fig7}(c)] is proportional to $\gamma_r/(\gamma_r+\gamma_{nr})$, where $\gamma_r$ is the inverse of the radiative lifetime, and $\gamma_{nr}$ the inverse of the non-radiative one. On the other hand, the inverse of the PL decay time [Fig.\ref{fig7}(d)] is equal to $\gamma_r+\gamma_{nr}$. Consequently, assuming that $\gamma_r\gg\gamma_{nr}$ at zero temperature, the time-integrated PL signal intensity can be rewritten as:
\begin{equation}
I(T)=I(0)\frac{\gamma_r}{\gamma_r+\gamma_{nr}}
\label{eq7mod}
\end{equation}

so that the ratio of the time-integrated PL signal intensity with the PL decay time provides the temperature dependence of the radiative rate $\gamma_r$. This quantity is plotted in the inset of Fig.\ref{fig7}(d). Within our experimental error, we observe a temperature-independent value, as expected for zero-dimensional systems. Such a behavior could be anticipated from the identical values of the activation energies $E_{a}^{(1)}$ and $E_{a}^{(2)}$, any temperature dependence of the radiative rate inducing different values for $E_{a}^{(1)}$ and $E_{a}^{(2)}$.

Therefore, we conclude that the radiative lifetime is constant as a function of temperature, and that the fast recombination dynamics at high temperature is solely due to non-radiative recombination, responsible for the emission decrease on raising the temperature [Fig.\ref{fig7}].

\section{CONCLUSION}\label{sec:conclu}

We revisited the fundamental opto-electronic properties of G-centers in silicon in order to complement the literature collected earlier until the late eighties. We characterized the saturation power by means of ensemble measurements displaying a sublinear increase as a function of incident power. We unraveled the recombination dynamics, occuring on a 6 ns time-scale at low temperature, without any significant variations as a function of the proton irradiation fluence. We quantitatively interpreted the vibronic spectrum by non-perturbative calculations of the acoustic phonon sideband, leading to an estimation of 1.6$\pm$0.1 $\angstrom$ for the spatial extension of the electronic wave function in a G-center. Finally, we recorded the temperature dependence of the emission spectrum and recombination dynamics, and we demonstrated that the radiative lifetime is constant as a function of temperature. Given the tremendous potential for manipulating and controlling point defects hosted in a silicon matrix and emitting in the telecommunications wavelength range, we believe that our optical characterizations of G-centers in silicon will stimulate further experiments and contribute to the expansion of this new field of research in quantum technologies.

\textbf{Acknowledgments}

We gratefully acknowledge C. L'Henoret for his technical support at the mechanics workshop, A. Dr\'{e}au, I. Philip, P. Valvin and B. Gil for helpful discussions. This work was financially supported by the network ULYSSES (ANR-15-CE24-0027-01) funded by the French ANR agency and the German DFG (PE 2508/1-1). C.V. and G.C. acknowledge the Institut Universitaire de France. A.K. acknowledges financial support from the Research Council of Norway via MIDAS project.

$^\dag$e-mail: marco.abbarchi@im2np.fr \\
$^\ast$e-mail: guillaume.cassabois@umontpellier.fr
 


\bibliographystyle{apsrev4-1}
\bibliography{biblio}

\end{document}